# Evolution of Pulse width modulation techniques of modular multilevel converter

Khushbu Saradva[1], Meeta Matnani[2], Tapankumar Trivedi[3]
[1]Student Member, IEEE

*Abstract-* **The modular multilevel converter (MMC) is a promising converter technology for various high-voltage high-power applications. The reason to that is low-distortion output quantities can be achieved with low average switching frequencies per switch and without output filters. Modular multilevel converter pulse width modulation based control approaches are evaluated. The multicarrier PWM techniques such as Phase disposition PWM (PDPWM), Phase opposition disposition PWM (PODPWM), Alternate phase opposition disposition PWM (APODPWM) is employed and a comparative study is done based on the quality of the load voltage and load current output. This paper presents the simulation of single phase modular multilevel converter based on Half-Bridge sub module topology. Simulation has been carried out for various modulation techniques using MATLAB/Simulink and the results are observed.**

INTRODUCTION

Modular multilevel converters have great potential in high-power applications, such as DC interconnections, DC power grids, and off-shore wind power generation are in need of accurate power flow control and high efficiency power conversion in order to reduce both their operating costs and their environmental impact [1]. High power converters for utility applications require line-frequency transformers for the purpose of enhancing their voltage or current rating. The use of line-frequency transformers, however, not only makes the converter heavy and bulky, but also induces the so-called DC magnetic flux deviation when a single-line-to-ground fault occurs [2].
MMC applications are mainly in the field of high-voltage high power applications such as in flexible A.C transmission system (FACTS) applications, small motor drives such as laminators, compressors, mills and large electric drives systems. Due to its power electronic switches balancing and distributed voltage stress, multilevel nature which means having nearly stepped sinusoidal waveforms and reduced harmonics. Its working principle is based on producing small output voltage steps which results in better power quality. They operate at low voltage levels and also at a low switching frequency so that the switching losses are also reduced. Other advantages include better electromagnetic compatibility due to the low dv/dt transitions. Generally multi-level inverters employ a number of cascaded power switches in order to increase either the voltage or power capability of inverter, the idea behind this concept is to use multiple small voltage level sources including storage devices , such as capacitors or batteries to generate a required signal taking advantage of using lower-rated semiconductors.

STRUCTURE OF MMC

Modular multilevel converter (MMC) topology was firstly proposed by Lesnicar and Marquardt in 2003. The main circuit a three-phase MMC is outlined in Figure 1. Each of the three parallel-connected phase legs comprises two phase arms. Each arm consists of a string of series-connected submodules, and a series inductor. Each submodule contains a switch-mode half-bridge and a capacitor [4]. A specific submodule can be inserted or bypassed in this series-connection by control of the associated half-bridge. The desired instantaneous voltage across the arm is controlled by series connecting a suitable number of charged sub module capacitors in the arm as commanded by the modulator.
When modeling a converter arm, there is some resistive voltage drop that appears across conducting switches, cables, connections and the inductor, which is often disregarded. This is reasonable, as in a large-scale converter resistive voltage drops need to be as small as possible, because they cause losses. Dimensioning an MMC is not an obvious procedure. It depends very much on each specific application. The first step is to decide the number of sub modules per arm, depending on the voltage level the converter will be connected to and the ratings of the semiconductors that are more cost efficient[6]. The number of levels determines also the nominal capacitor voltage in each sub module. The arm inductor values are determined based on external fault criteria, and should be high enough to limit the current from an AC short-circuit. On the other hand, from a control perspective, large inductors will slow down the effect of the controllers, which is undesirable. The fault currents also contribute to the capacitor dimensioning, where the output power level, as well as the level of active control applied are also important factors.

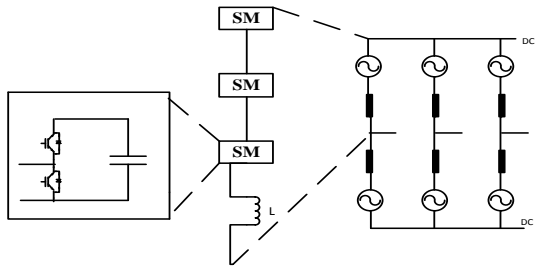

Figure 1 Schematic of a three phase MMC

Figure 2 shows the single-phase equivalent circuit of the MMC. $L_0$ and $R_0$ are the arm inductance and equivalent arm resistance,

respectively. $U_{dc}$ and $I_{dc}$ are the total dc bus voltage and dc current, respectively. $u_{vj}$ is the converter output voltage of phase j at point V whereas $i_{vj}$ is the corresponding line current. The arm voltages generated by the cascaded SMs are expressed as $u_{pj}$ and $u_{nj}$ where the subscripts p and n denote the upper (positive) and lower (negative) arms, respectively. According to Figure 2 and [11], the corresponding arm currents can be expressed as

$$i_{pj} = i_{diffj} + \frac{i_{vj}}{2} \ldots\ldots\ldots\ldots(1)$$

$$i_{nj} = i_{diffj} - \frac{i_{vj}}{2} \ldots\ldots\ldots\ldots(2)$$

Where $i_{diffj}$ is the inner difference current of phase j, which flows through both the upper and lower arms and is given as

$$i_{diffj} = \frac{i_{pj} + i_{nj}}{2} \ldots\ldots\ldots\ldots(3)$$

According to [11], the MMC can be characterized by the following equations:

$$u_{vj} = e_j - \frac{R_0 i_{vj}}{2} - \frac{L_0}{2} \cdot \frac{di_{vj}}{dt} \ldots\ldots\ldots(4)$$

$$L_o(\frac{di_{diffj}}{dt}) + R_o i_{diffj} = \frac{U_{dc}}{2} - \frac{u_{pj} + u_{nj}}{2} \ldots\ldots(5)$$

$here, (j = a,b,c)$

Where $e_j$ in (4) is the inner emf generated in phase j and is expressed as

$$e_j = \frac{u_{nj} + u_{pj}}{2} \ldots\ldots\ldots(6)$$

According to (4), when $u_{vj}$ is regarded as the ac network voltage, the current $i_{vj}$ can be controlled directly by regulating the control variable $e_j$.

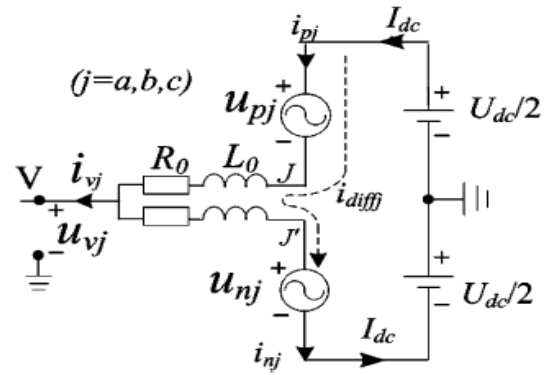

Figure 2 Single-phase equivalent circuit of the MMC

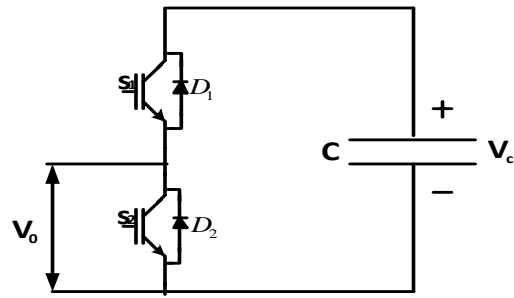

Figure 3 Configuration of Half bridge sub-module
Table 1 switching state of Half-bridge sub module

| State | $S_1$ | $S_2$ | $V_0$ |
|-------|-------|-------|-------|
| 1 | ON | OFF | $V_c$ |
| 2 | OFF | ON | 0 |

OPERATION OF HALF BRIDGE SM

The sub module produces the required AC phase voltage. AC side terminals of the converter branch out between the two inductors of upper and lower arms. The submodule topology, being half-bridge circuit as shown in Figure 3, is composed of only two switches in cascade and a submodule capacitor paralleled to the switches. This simple and efficient structure allows the half-bridge circuit to dominate the others as the common submodule structure.

At any time, only one of the switches of half-bridge circuit should be ON. If $S_1$ is ON and $S_2$ is OFF, then the half-bridge circuit is "switched on" or "inserted to the current path". Else if $S_1$ is OFF and $S_2$ is ON, then the half-bridge circuit is "switched off" or "bypassed". The terminal voltage of half-bridge circuit is equal to the voltage across the submodule capacitor, Vc, if switched on/inserted or zero if switched off/bypassed. If both of the switches are ON then the submodule capacitor is short-circuited. If both of the switches are OFF then terminal voltage of the submodule is undetermined and according to the direction of the current, different voltages may appear at the terminals. Depending on the states of half-bridge circuit and the direction of submodule current, submodule capacitor is either charged or discharged.

MULTI CARRIER PWM STRATEGIES FOR MMC

Multi-Carrier PWM strategies is widely used, because it can be easily implemented to low voltage modules.. The level shifted PWM (LS-PWM) are In Phase Disposition (PD), Phase Opposition Disposition (POD) and Alternative Phase Opposition Disposition (APOD). In general the amplitude modulation index ($m_a$) for level shifted PWM technique is defined as the ratio of amplitude of the reference sine wave ($A_r$) to the amplitude of the carrier wave ($A_c$) and it is given in equation (1).

$$m_a = \frac{2A_r}{(m-1)A_c} \quad (1)$$

The frequency modulation index ($m_f$) for level shifted and phase shifted PWM techniques is defined as the ratio of frequency of the carrier wave ($f_c$) to the frequency of the reference sine wave ($f_r$) and it is given in equation (2).

$$m_f = \frac{f_c}{f_r} \quad (2)$$

A. Phase Disposition PWM (PDPWM)

PD method should be implemented with two different carrier sets for upper and lower arms in order to construct the phase voltage in N+1 level. The carrier set has again N identical carriers with amplitude of $V_{dc}/N$ and displaced contiguously in the $V_{dc}$ band, ranging from 0 to $V_{dc}$, Submodules in upper and lower arms are switched with the first and second carrier sets.

B. Phase Opposition Disposition PWM (PODPWM)

POD method should be implemented with a single carrier set for both upper and lower arms in order to construct the phase voltage in N+1 level. Submodules in upper and lower arms are switched with this carrier set. Phase Opposition Disposition (POD) PWM where the carriers above the zero reference are in phase but shifted by 180° from those carriers below the zero reference.

C. Alternate Phase Opposition Disposition PWM (APODPWM)

In order to construct the phase voltage in N+1 level, APOD method should be implemented with a single carrier set for both upper and lower arms. Alternative Phase Opposition Disposition (APOD) PWM where each carrier band is shifted by 180° from the adjacent band. also all the carriers have the same frequency and the adjustable amplitude (different or unequal Amplitudes).

SIMULATION RESULTS

In order to verify the modulation method, a three phase MMC is simulated using Matlab/Simulink and its performance is studied for RL load.

Table 2 simulation parameters

| DC link voltage | $100V$ |
|---|---|

| Carrier frequency | $1 kHz$ |
|---|---|
| Reference frequency | $50 Hz$ |
| Dc link capacitance of SM | $1100 \mu F$ |
| Arm inductance | $20 mH$ |
| Frequency Index | 20 |
| Modulation Index | 0.9 |
| Load | $10 ohm, 30 mH$ |

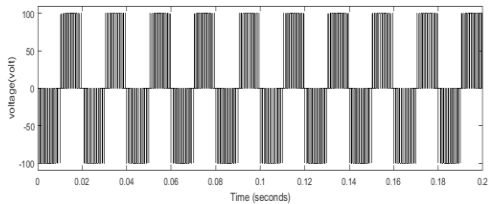

Figure 4 Output voltage for 3 level MMC with PD modulation technique

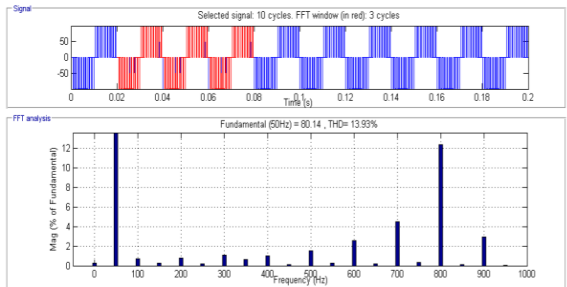

Figure 5 FFT analysis &THD for 3 level MMC with PD modulation technique

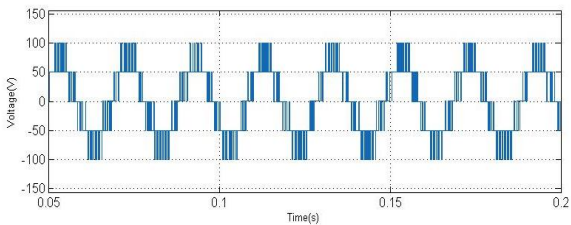

Figure 6 Output voltage for 5 level MMC with PD modulation technique

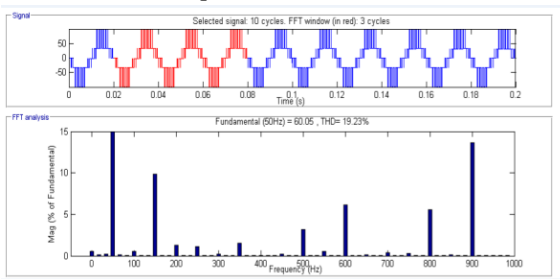

Figure 7 FFT analysis & THD for 5 level MMC With PD modulation rechnique

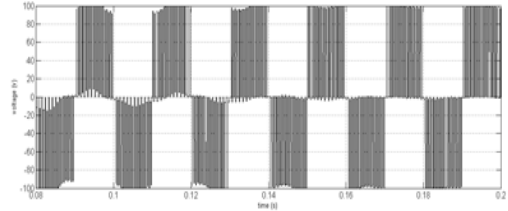

Figure 8 Output voltage for 3 level MMC with PSC modulation technique

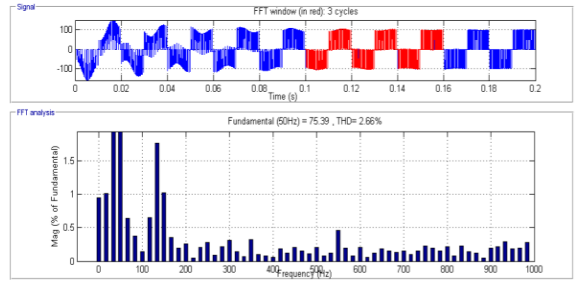

Figure 9 FFT analysis &THD for 3 level MMC with PSC modulation technique

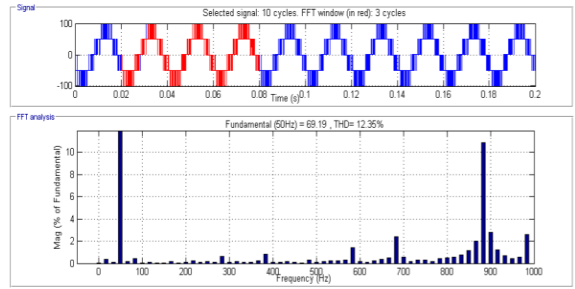

Figure 10 FFT analysis &THD for MMC with PD modulation technique

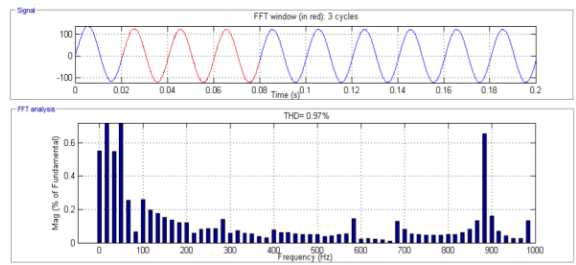

Figure 11 FFT analysis & current THD for MMC with PD modulation technique

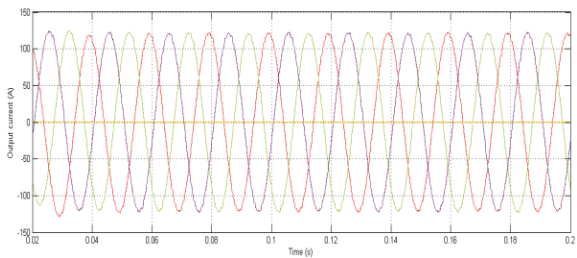

Figure 12 Output current for MMC with PD modulation technique

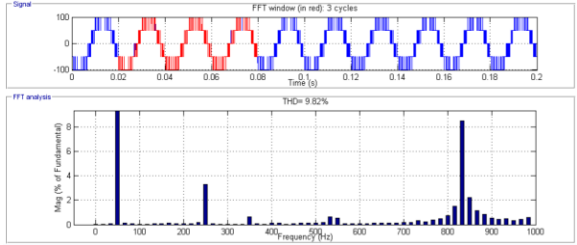

Figure 23 FFT analysis &THD for MMC with POD modulation technique

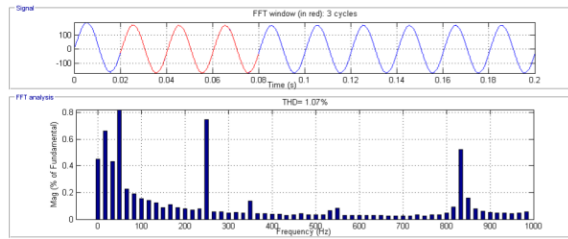

Figure 14 FFT analysis & current THD for MMC with POD modulation technique

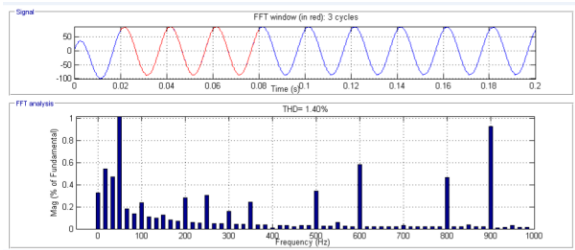

Figure 15 FFT analysis & current THD for 5 level MMC with PD modulation technique

## CONCLUSION

In this paper, single phase three-level modular multilevel inverter based on Half bridge cells configuration employing PDPWM,PODPWM,PSC strategies with RL load is simulated using Matlab/Simulink software. From the work done it is concluded that THD of output current varies from 0.97% to 1.40% for modular multilevel converter, using PD, POD and PSC modulation technique A fundamental frequency modulation techniques schemes such optimized harmonic stepped waveform and selective harmonic elimination could be used for further researches.